\documentclass[trackchanges, twocolumn]{aastex701}
\usepackage{amsmath}
\usepackage{amssymb}
\usepackage{amsfonts}
\usepackage{times}
\usepackage{color}
\usepackage{subcaption}

\renewcommand{\d}{\mathrm{d}}

\begin{document}

\title{Anchoring the Universe with Characteristic Redshifts using \\ Raychaudhuri Equation Informed Reconstruction Algorithm (REIRA)} 

\correspondingauthor{Shibendu Gupta Choudhury} 

\author[orcid=0000-0002-3840-9456,sname='Gupta Choudhury']{Shibendu Gupta Choudhury}
\email[show]{pdf.schoudhury@jmi.ac.in}
\affiliation{Centre for Theoretical Physics, Jamia Millia Islamia, New Delhi-110025, India}
 
\author[orcid=0000-0002-2701-5654,sname='Mukherjee']{Purba Mukherjee} 
\email[show]{pdf.pmukherjee@jmi.ac.in}
\affiliation{Centre for Theoretical Physics, Jamia Millia Islamia, New Delhi-110025, India}

\author[orcid=0000-0001-9615-4909,sname='Sen']{Anjan Ananda Sen}
\email[show]{aasen@jmi.ac.in}
\affiliation{Centre for Theoretical Physics, Jamia Millia Islamia, New Delhi-110025, India}

\begin{abstract}
We study the robustness and physical implications of a set of \emph{characteristic redshifts} that capture key features of the late-time Universe. Using both model-independent reconstructions as well as different dark energy (DE) parameterizations, we show that these redshifts remain stable across cosmological models and reconstruction algorithm, making them reliable geometric anchors of the expansion history. Moreover, the Alcock–Paczyński corrections at these redshift anchors are found to be unity with high statistical significance, making them natural isotropy points in the comoving distance–redshift relation. We also find that certain redshifts anchors $(z < 1)$ coincide with epochs where strong deviations from the Planck $\Lambda$CDM baseline are apparent irrespective of DE parametrisation like CPL or reconstruction algorithm, indicating their potential as probes of new physics in cosmological evolution. Finally, we demonstrate, for the first time, that a \emph{Raychaudhuri Equation Informed Reconstruction Algorithm},   substantially enhances the precision of the inferred distance measures and the Hubble expansion rate as well as results tighter constraints in the DE parameter space. These results demonstrate that combining geometric reconstruction with physics-informed kinematic information offers a powerful and consistent algorithm to probe new physics in the late-time dynamics of our Universe.
\end{abstract}



\section{Introduction}

One of the most challenging questions in modern cosmology today is whether the concordance 
$\Lambda$CDM model provides a complete description of the Universe, 
or whether new physics in the DE sector is required. 
Several observations already hint at the latter, including the Hubble tension, 
the $S_{8}/\sigma_{8}$ tension \citep{Riess:2021jrx, Riess:2022mme, DiValentino:2021izs, Abdalla:2022yfr, CosmoVerseNetwork:2025alb}, the abundance of unexpectedly massive galaxies 
at $z > 10$ observed by James Webb Space Telescope (JWST) \citep{Haslbauer2022, Lovell2023, Boylan2023}, and the recent baryon acoustic oscillations (BAO) results from Dark Energy Spectroscopic Instrument public data release 2 (DESI DR2) \citep{DESI:2024mwx, DESI:2025zgx}
combined with Type Ia supernova (SnIa) \citep{Scolnic:2021amr, Rubin:2023jdq, DES:2024jxu} and cosmic microwave background (CMB) \citep{Planck:2018vyg, Tristram:2023haj, ACT:2025fju} measurements. With upcoming data 
from LSST \citep{LSST:2008ijt, 2009arXiv0912.0201L, 2022arXiv220802781B} and Euclid \citep{Euclid:2019clj, Euclid:2021xmh, Euclid:2024yrr, EuclidTheoryWorkingGroup:2012gxx}, resolving these issues is a pressing goal.

While most studies probe DE using specific scalar-field models 
(both non interacting or interacting with dark matter) \citep{Amendola_Tsujikawa_2010, CosmoVerseNetwork:2025alb, Park:2025fbl, Wolf:2025acj,Toomey:2025xyo,Hossain:2025grx} 
or parametric forms of its equation of state $(w_{\rm DE})$ \citep{Chevallier:2000qy, Linder:2002et, Linder:2005in, Jassal:2005qc,Sohail:2024oki,Guedezounme:2025wav,Shah:2025ayl,RoyChoudhury:2025dhe}, energy density $(\rho_{\rm DE})$ \citep{DiValentino:2020naf, Dinda:2025iaq, Adil:2023exv, Cheng:2025lod}, pressure ($p_\mathrm{DE}$) \citep{Sen:2007gk, Cheng:2025lod} etc., a more direct, 
model-independent approach is to reconstruct the comoving distance $D_{M}$ 
and its derivative from data, thereby obtaining $H(z)$ without assuming a DE model 
[See also parametrisation for $H(z)$ \citep{Capozziello:2018jya, Dutta:2018vmq,Roy:2022fif}, scale factor $a(t)$ \citep{Mukhopadhyay:2024fch, Choudhury:2025bnx}, reconstruction for $w_{\rm DE}$ as well $\rho_{\rm DE}$ \citep{Dutta:2019pio, Berti:2025phi, Dinda:2024ktd, Jiang:2024xnu, Mukherjee:2024ryz, DESI:2025wyn}]. 
\citet{Mukherjee:2025ytj} recently applied this method to BAO and SnIa datasets, 
anchored by Planck-2018 \citep{Planck:2018vyg} $\Lambda$CDM early-universe physics. They identified 
seven characteristic redshifts and measured $H(z)$, finding \textbf{significant 
deviations from Planck-$\Lambda$CDM predictions at three redshifts for $z < 1.0$, 
while higher-redshift results remain consistent with Planck-$\Lambda$CDM behaviour}. This points to possible new physics 
in the low-redshift Universe, while the $\Lambda$CDM framework holds at earlier times.

These findings highlight the importance of low-redshift precision cosmology.
With forthcoming datasets from LSST and Euclid, it will be crucial in determining whether 
these anomalies confirm a breakdown of $\Lambda$CDM and unveil entirely new 
physics in the DE sector.

In this work, we examine the robustness of the characteristic redshifts identified by \citet{Mukherjee:2025ytj}, obtained using the geometric properties of cosmological distances, as anchors of cosmic evolution. We show that these redshifts are not only robust under different DE parameterizations but also directly linked to observable signatures, such as the Alcock–Paczyński (AP) effect \citep{Alcock:1979mp} in BAO data. Furthermore, we demonstrate, \emph{for the first time}, that \textbf{incorporating kinematic information through the Raychaudhuri Equation (RE)} \citep{Raychaudhuri:1953yv, Ehlers1993} into the cosmological reconstruction framework leads to substantial improvements. In particular, the inclusion of such physically motivated constraints sharpens the reconstruction of distance measures, yields a more accurate determination of the expansion history 
$H(z)$ and provides tighter bounds on the DE parameter space. Together, these results highlight the utility of combining purely geometric reconstructions with kinematic information to achieve a more robust and physically consistent understanding of cosmic acceleration.

This paper is organized as follows. In Section~\ref{sec:charz}, we introduce the concept of \emph{Characteristic Redshifts} derived from cosmic distance measures. Section~\ref{sec:data} describes the observational datasets used in our analysis. In Section~\ref{robust}, we test the robustness of the derived characteristic redshifts. Section~\ref{sec:ap} explores their physical significance through the Alcock--Paczy\'nski (AP) correction. In Section~\ref{sec:markers}, we discuss how characteristic redshifts may serve as markers for new physics. Section~\ref{sec:rc} presents a physics-informed reconstruction approach based on the RE. Section~\ref{sec:like} details the construction of a compressed likelihood from the reconstructed data vectors. Finally, we summarize our conclusions in Section~\ref{sec:conclusion}.

\section{CHARACTERISTIC REDSHIFTS FROM COSMIC DISTANCES} \label{sec:charz}

In practice, cosmology does not provide direct access to physical distances. Instead, distances are inferred from observables such as redshift, angular size, and flux. This motivates the introduction of three key distance measures \citep{Hogg:1999ad}: the comoving distance $d_M(z)$, the angular diameter distance $d_A(z)$,  and the luminosity distance $d_L(z)$, given by,
\begin{equation}\label{disrel}
\begin{split}
    d_M(z) &\equiv c \int_{t}^{t_0} \frac{\d t}{a(t)} 
    = c \int_0^z \frac{\d z}{H(z)} , \\     d_A(z) &= \frac{d_M(z)}{1+z}, \\
    d_L(z) &= (1+z)\, d_M(z),
     \end{split}
\end{equation}
where $a(t)$ is the scale factor of the Universe, related to the redshift $z$ as $a=(1+z)^{-1}$, $H(z)$ is the Hubble parameter and $c$ is the speed of light. We can define dimensionless measures for these distances as $D_M(z)\equiv H_0 d_M(z)/c$, $D_A(z)\equiv H_0 d_A(z)/c$, and $D_L(z)\equiv H_0 d_L(z)/c$, where $H_0$ is the present day ($t=t_0$) value of the Hubble parameter.

In a spatially flat Friedmann--Lemaître--Robertson--Walker (FLRW)
geometry, all cosmological distance measures are determined solely by the expansion history, encoded in the Hubble parameter $H(z)$. Beyond their basic definitions, the derivatives of these distances with respect to redshift carry additional geometric information. In a spatially flat FLRW Universe, the simplicity of the distance relations makes it possible to identify characteristic redshifts at which certain derivative-based conditions are satisfied. In \citet{Mukherjee:2025ytj}, a set of $7$ redshifts $z_i$ (for $i=1, 2,..., 7$), referred to as \emph{characteristic redshifts}, is introduced and defined through the following relations:
\begin{equation}\label{rel}
    D_{\{X\}}  (z_i)=D_{\{Y\}}^\prime(z_i), ~~  \left(\{X\} ~\text{and}~ \{Y\} \equiv M, A~ \text{or}~ L\right).
\end{equation}

To determine the characteristic redshifts from the conditions above, one can evaluate the cosmological distances for a given expansion history or reconstruct these distances directly from the observational data. The latter analysis using non-parametric Gaussian processes and free-form knot reconstruction has already been reported in \citet{Mukherjee:2025ytj}. To compare different DE models through the lens of these characteristic redshifts, it is convenient to derive general evolution equations for these distances, which can then be solved directly choosing different DE behavior. Since the distance measures are related with each other through Eq. \eqref{disrel}, it is sufficient to obtain the evolution equation for any one of these distances. We derive such an evolution equation for $D_A$ in this section. Assuming General Relativity as the underlying theory of gravity, this evolution equation at late times comes out as,
\begin{equation}\label{angd}
   x^{\prime\prime} + \frac{3x^\prime}{2(1+z)} 
   \left[ 1 + w_\mathrm{DE}(z) \left\lbrace 1 - \Omega_{m0}(1+z)^3 x^{\prime 2} \right\rbrace \right] = 0,
\end{equation}
where $x\equiv (1+z)D_A$,  $w_\mathrm{DE} = p_\mathrm{DE}/\rho_\mathrm{DE}$ is the DE equation of state ($p_\mathrm{DE}$ is the pressure and $\rho_{\rm DE}$ is the energy density in normalized units), $\Omega_{m0}$ is the matter density parameter at current times, and prime denotes differentiation with respect to redshift. Using different assumptions for the DE equation of state corresponding to different models, we can solve Eq. \eqref{angd} and find the corresponding characteristic redshifts. In section \ref{robust}, we undertake such an analysis to establish the robustness of these redshifts.

\section{Observational datasets} \label{sec:data}

We will primarily use two key datasets in our analysis:  BAO distance measurements from DESI DR2 \citep{DESI:2025zgx} and SnIa dataset from the five-year observation campaign of the Dark Energy Survey (DES 5YR) \citep{DES:2025xii}. In addition to the above two low redshift measurements, we incorporate constraints from the CMB via the so-called shift parameters, which include the CMB shift $R$, the acoustic angular scale $l_a$, and the current physical baryon density \( \omega_b \). These parameters provide a compact summary of CMB data \citep{Wang:2007mza, Bansal:2025ipo}, assuming standard $\Lambda$CDM physics prior to recombination, and are used in our analysis accordingly.

\begin{figure*}[htb]
    \centering
    \includegraphics[width=0.2325\linewidth]{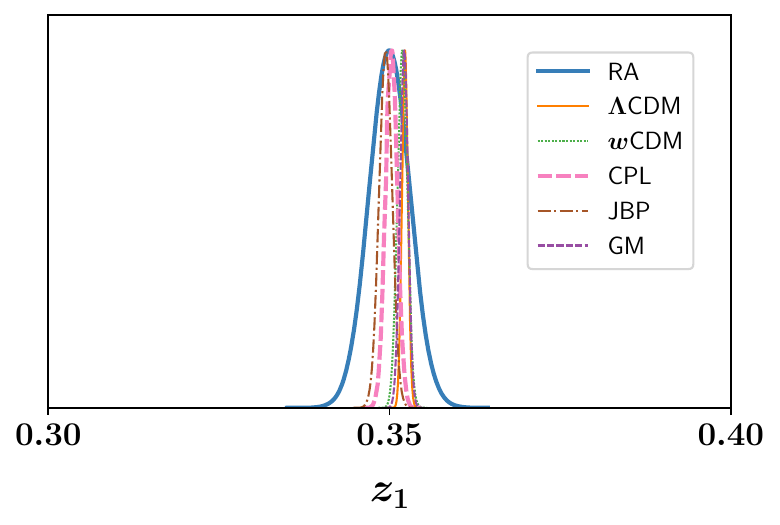}\includegraphics[width=0.22\linewidth]{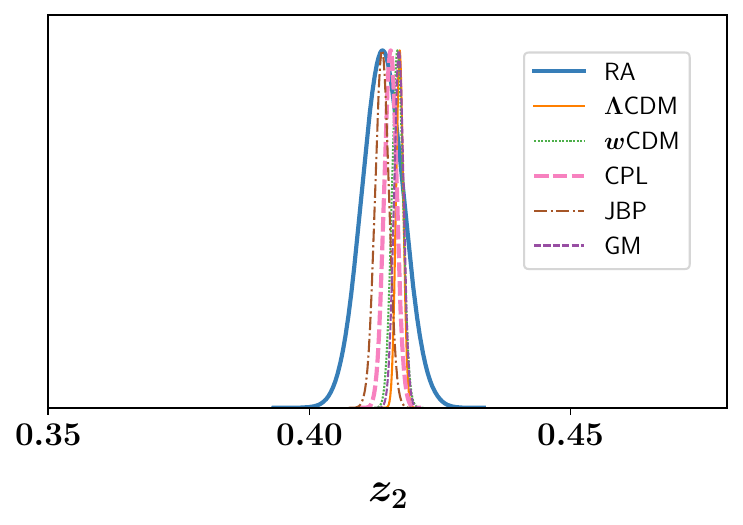}  \includegraphics[width=0.215\linewidth]{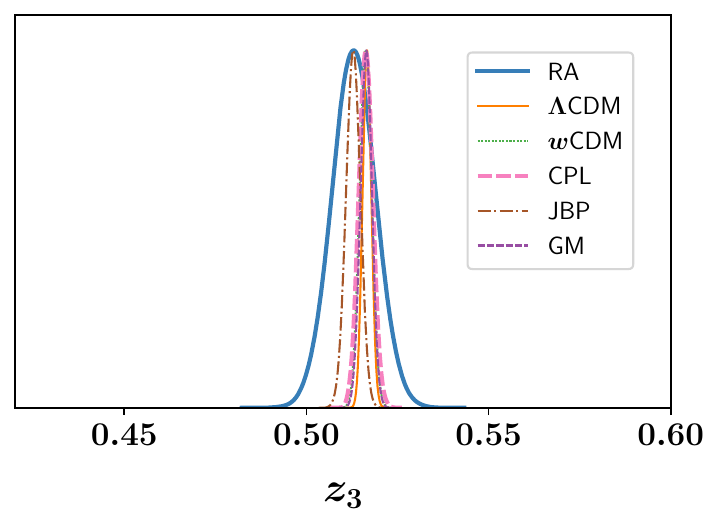}\includegraphics[width=0.2255\linewidth]{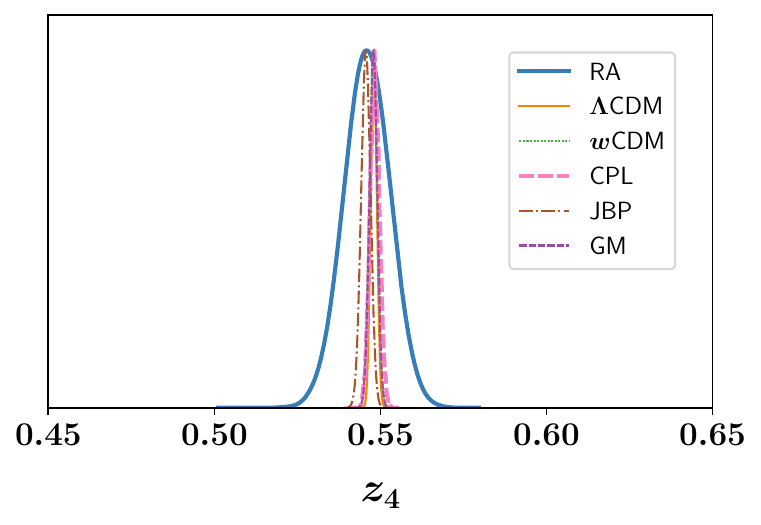} \includegraphics[width=0.22\linewidth]{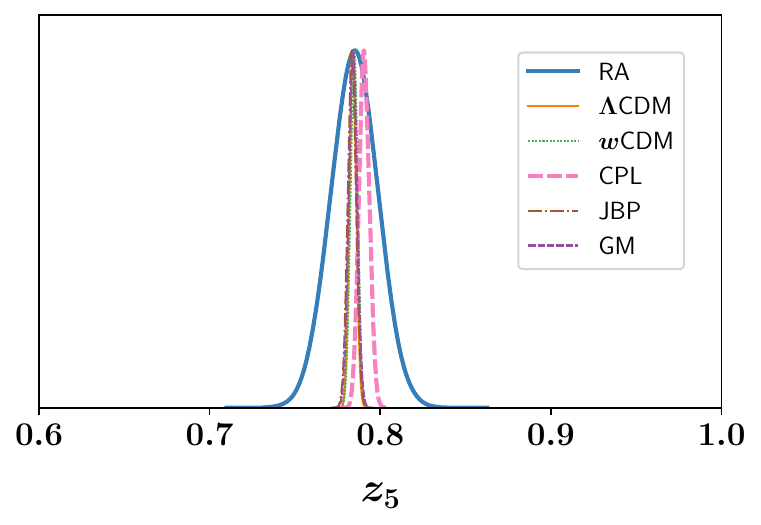}\includegraphics[width=0.215\linewidth]{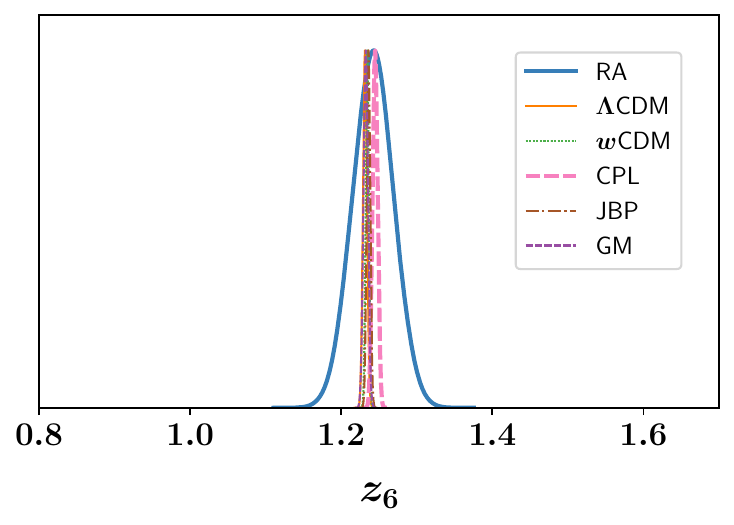}~~\includegraphics[width=0.215\linewidth]{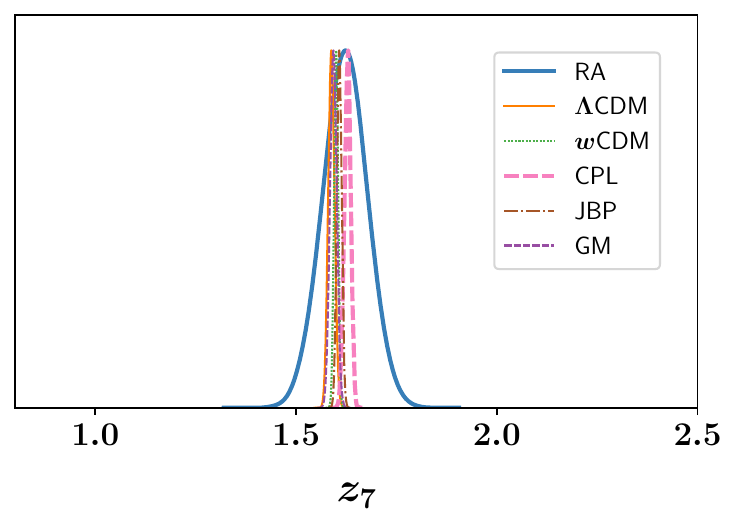}
    \caption{Characteristic redshifts and their uncertainty for RA and different DE parameterizations.}\label{fig_zs}
\end{figure*}

\section{Robustness of Characteristic Redshifts}\label{robust}

To establish the robustness of the characteristic redshifts $z_i$ across different DE parametrization, we consider the following models: $\Lambda$CDM, $w$CDM,  Chevallier-Polarski-Linder (CPL) parametrization \citep{Chevallier:2000qy, Linder:2002et, Linder:2005in}, Jassal-Bagla-Padmanabhan (JBP) parametrization \citep{Jassal:2005qc} and Scale factor parametrization \citep{Sen:2001xu, Mukhopadhyay:2024fch}.  We first constrain these models using the combined DESI DR2 + DES 5YR + CMB datasets through a Bayesian Markov Chain Monte Carlo (MCMC) analysis. 

We obtain the corresponding redshifts and their uncertainty directly from the MCMC chains by solving Eq. \eqref{angd}, except for the $\Lambda$CDM model. For $\Lambda$CDM model, we use the results of Planck 2018 \citep{Planck:2018vyg} to obtain the redshifts. To compare the redshifts for different models with those obtained directly from model independent reconstruction of \citet{Mukherjee:2025ytj}, we include the result for free-form knot-based reconstruction algorithm (RA) for spline order $k = 4$. The robustness of these redshifts is illustrated in Fig. \ref{fig_zs} which shows that these redshifts are remarkably stable and largely independent of different DE parameterizations. 

 \emph{This demonstrates that these characteristic redshifts are imprinted in the observational data from DESI DR2 for BAO measurments and DES 5YR for SnIa measurements, irrespective of DE models or model independent reconstructions.}

\section{Physical Significance of Characteristic Redshifts: minimal Alcock--Paczy\'nski (AP) Correction} \label{sec:ap}

The Alcock--Paczyński (AP) effect \citep{Alcock:1979mp} arises because the cosmological dependence of comoving distances differs along and across the line of sight. An intrinsically isotropic feature will appear isotropic in comoving coordinates only if the conversion from observed redshift and angle to distances is carried out with the correct ratio, $D_M^\prime(z)/D_M(z)$. At the characteristic redshifts, this ratio is constrained by derivative-based relations \eqref{rel}, implying that the AP correction factor reduces to unity, highlighting the robustness of these redshifts.

\begin{figure*}
    \centering
    \includegraphics[width=0.23\linewidth]{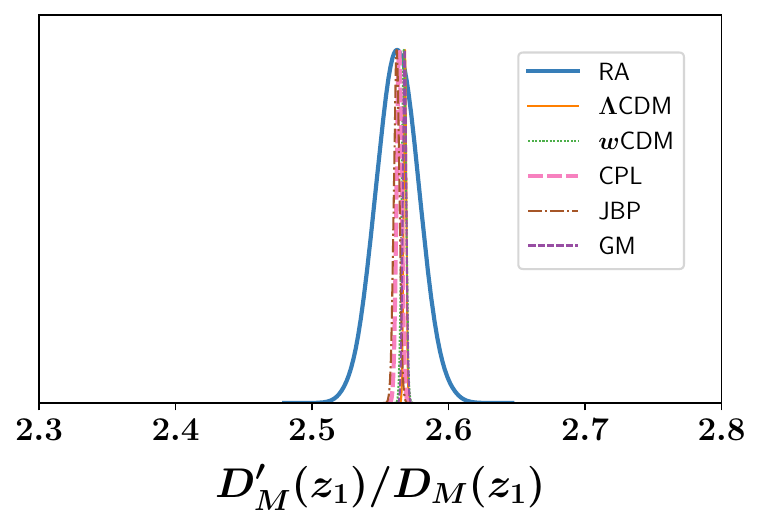} \includegraphics[width=0.23\linewidth]{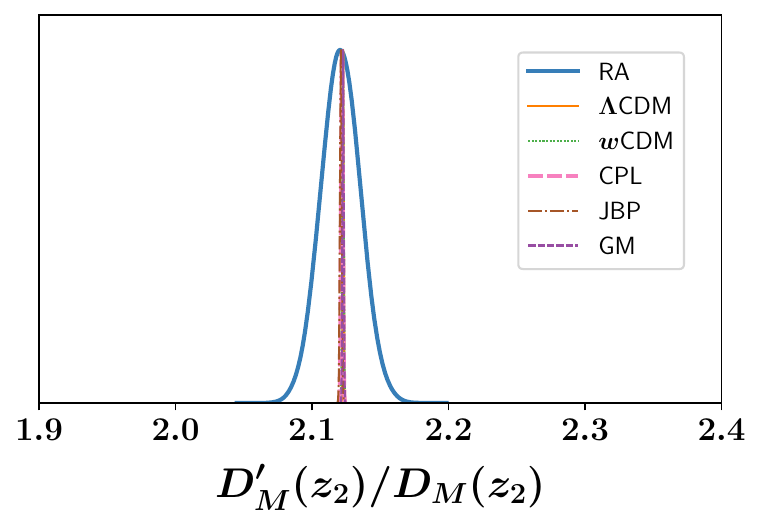}  \includegraphics[width=0.23\linewidth]{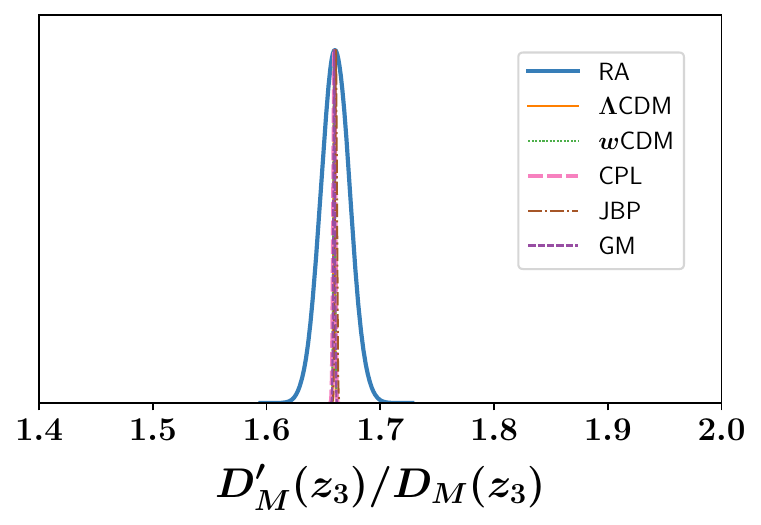} \includegraphics[width=0.2235\linewidth]{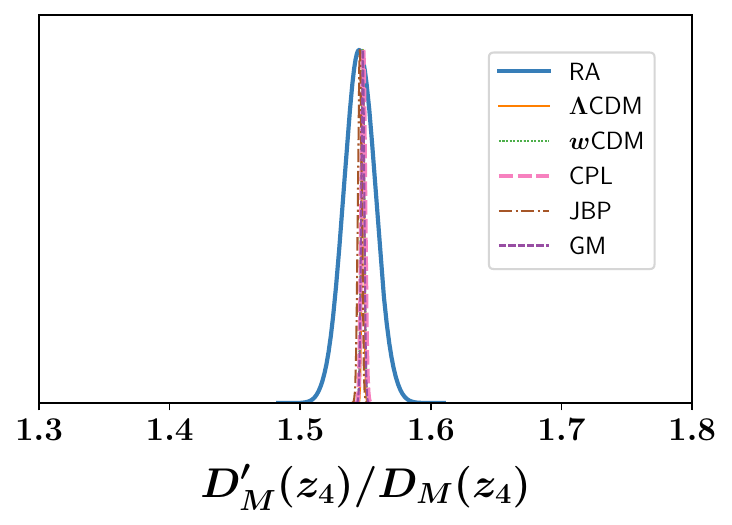} \includegraphics[width=0.23\linewidth]{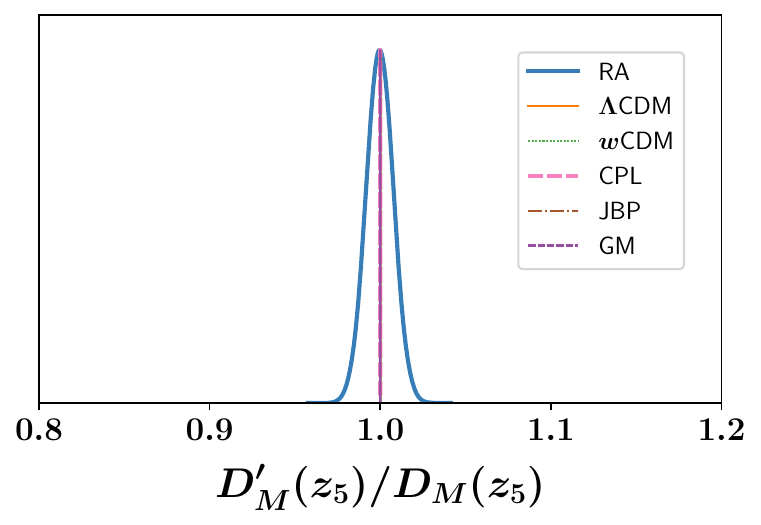} \includegraphics[width=0.23\linewidth]{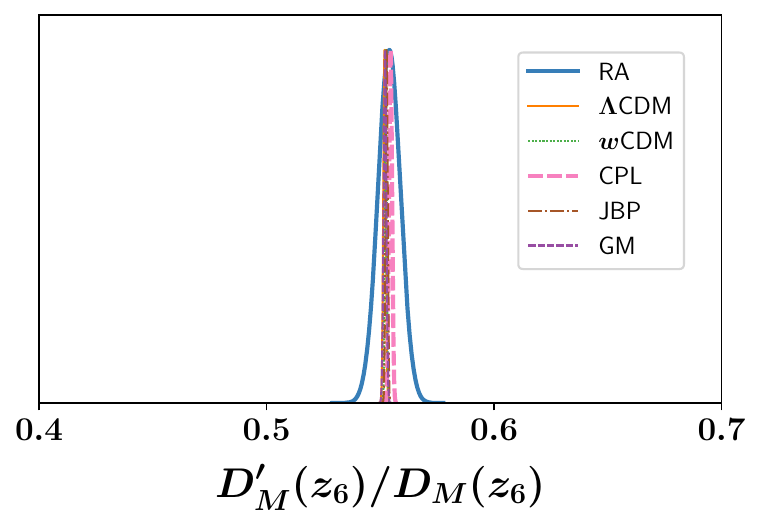}\includegraphics[width=0.23\linewidth]{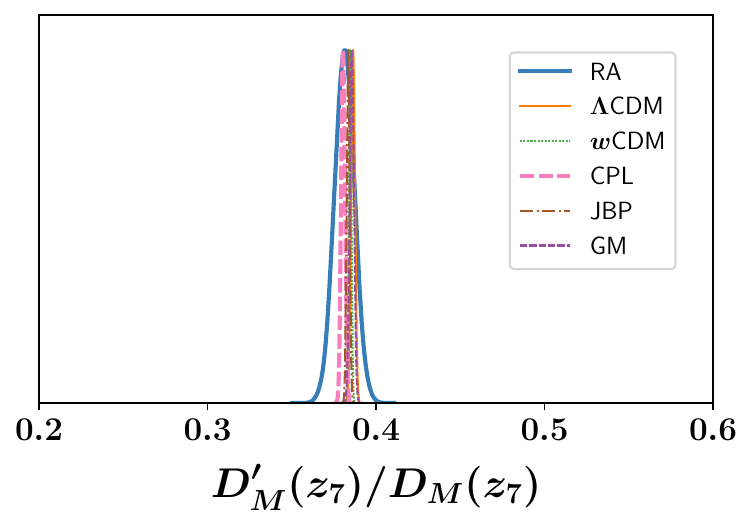}
    \caption{$D_M^\prime/D_M$ and its uncertainty at Characteristic redshifts for RA and different DE parameterizations.}\label{fig_APs}
\end{figure*}

In our analysis, we explicitly evaluate $D_M^\prime/D_M$ at these special redshifts for the DE parameterizations discussed above using the MCMC chains. We also perform the same evaluation using the RA. We find that the values are practically identical across different models and reconstruction techniques, as illustrated in Fig.~\ref{fig_APs}. Consequently, the AP correction factor effectively becomes unity at these redshifts for all scenarios considered.

This result provides crucial physical motivation for the significance of these redshifts: \emph{they serve as robust epochs in the cosmological evolution where model dependence is minimized and AP factor is unity, reinforcing their potential as powerful and reliable probes for exploring possible new physics.}

\section{Characteristic Redshifts: Markers for New Physics} \label{sec:markers}

Building on the robustness and physical significance of the characteristic redshifts, we now turn to their potential role in uncovering new physics beyond the $\Lambda$CDM paradigm. Our focus will be on reconstruction-based results, which we compare with the widely studied CPL parameterization.

\begin{figure*}[t]
    \centering
    \begin{subfigure}[b]{0.45\textwidth}
        \includegraphics[width=\textwidth]{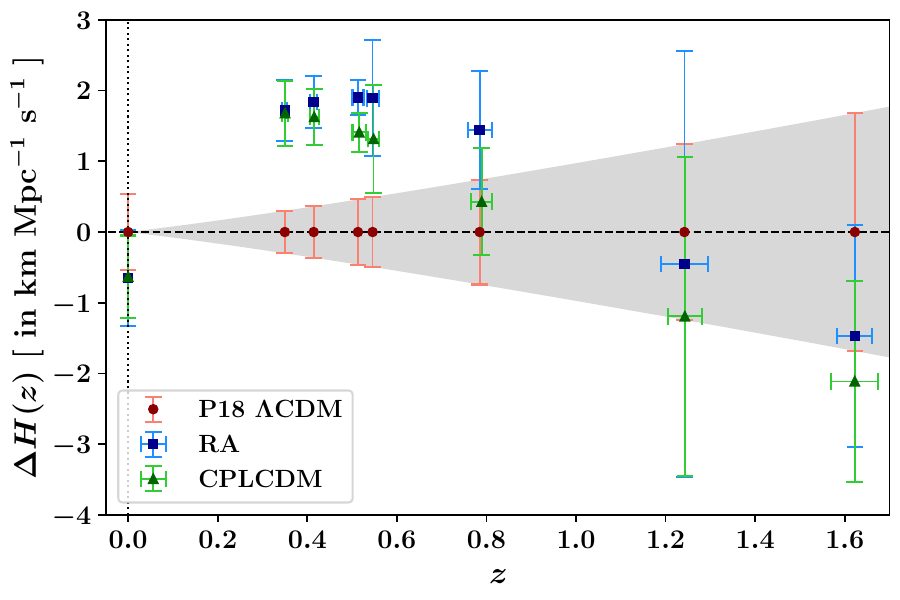}
        \caption{}
        \label{fig4:sub1}
    \end{subfigure}
       \begin{subfigure}[b]{0.45\textwidth}
        \includegraphics[width=\textwidth]{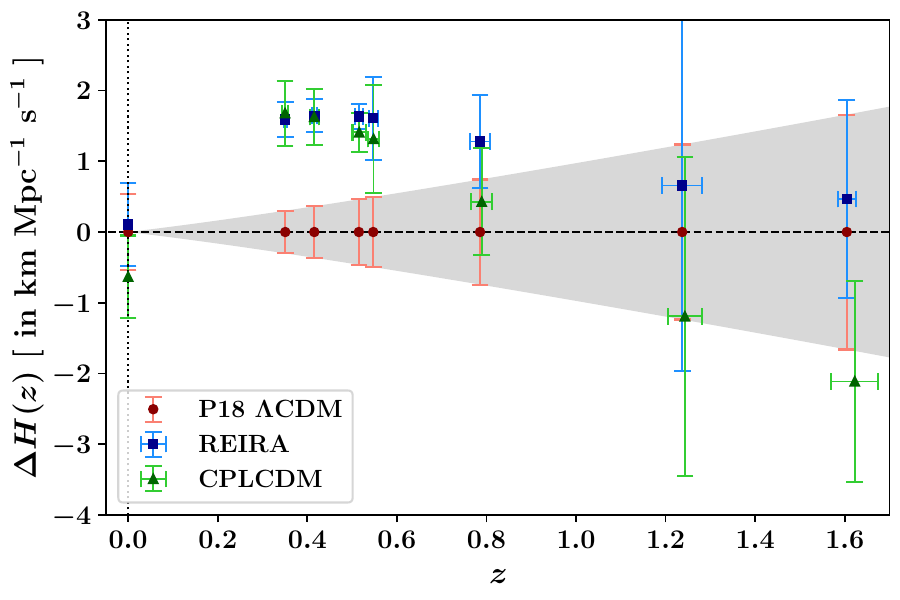}
        \caption{}
        \label{fig4:sub2}
    \end{subfigure}
       \caption{Deviation of the reconstructed Hubble parameter from the $\Lambda$CDM baseline ($\Delta H$) with 1$\sigma$ error-bars at the characteristic redshifts for RA and REIRA with a comparison to CPL shown in \ref{fig4:sub1} and \ref{fig4:sub2} respectively. The shaded region denotes consistency with the $\Lambda$CDM baseline.}
    \label{fig:overall}
\end{figure*}

Direct reconstruction of the Hubble parameter from observational data has revealed $4\text{--}5\sigma$ deviations from the Planck $\Lambda$CDM baseline in the redshift range $z \sim 0.35\text{--}0.55$ \citep{Mukherjee:2025ytj}. To probe this further, we plot in Fig.~\ref{fig4:sub1} the deviation of the Hubble parameter from the $\Lambda$CDM baseline,
\begin{equation}
    \Delta H(z_i) = H(z_i) - H_{\mathrm{P18}}(z_i),
\end{equation}
evaluated at the characteristic redshifts (including $z=0$) for both RA and the CPL model. For clarity, the shaded region denotes the range consistent with $\Lambda$CDM predictions. As shown, the \textbf{CPL values closely track those from RA, demonstrating that CPL can reproduce the reconstruction results with remarkable accuracy}. Notably, both exhibit large deviations in $\Delta H$ at $z_1$, $z_2$, and $z_3$. Specifically, the Hubble parameter shows $3-4\sigma$ tension for both RA as well as for CPL at $z_1$, $z_2$, and $z_3$\footnote{Tables \ref{tab:spline4} and \ref{tab:cpl} in Appendix further summarize the values of $D_A(z)$ and $E(z)=\frac{H(z)}{H_{0}}$ at the characteristic redshifts for RA, CPL, and baseline $\Lambda$CDM. They also highlight the statistical tension in these quantities.}. The findings establish these three characteristic redshifts as prominent cosmic markers, where deviations from Planck $\Lambda$CDM are most pronounced. 

\emph{This also suggests that the recently reported $3-4\sigma$ deviation from the $\Lambda$CDM behavior, obtained by combining DESI DR2 BAO data with various Type Ia supernova compilations and Planck-2018 CMB constraints using the CPL parametrization \citep{DESI:2025zgx}, may originate from the significant departure from $\Lambda$CDM, observed in Fig.~\ref{fig4:sub1} at these three low-redshift epochs identified in our study.}

\begin{figure*}
    \centering
    \includegraphics[width=0.23\linewidth]{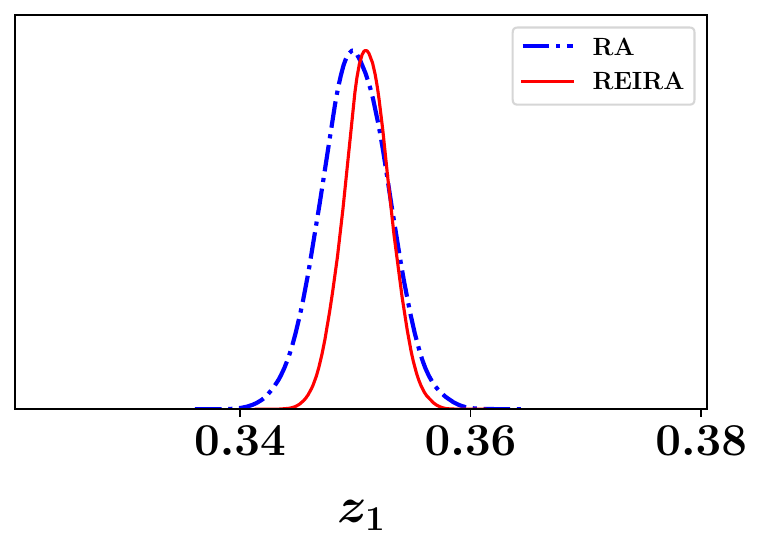}
    \includegraphics[width=0.23\linewidth]{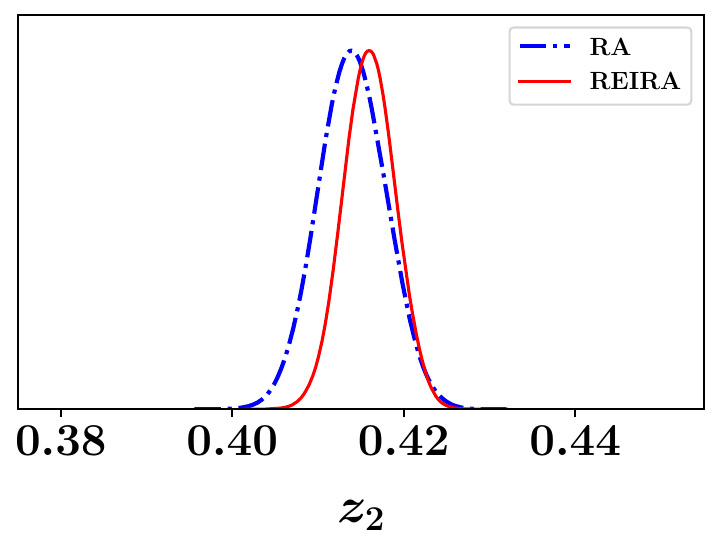}
    \includegraphics[width=0.23\linewidth]{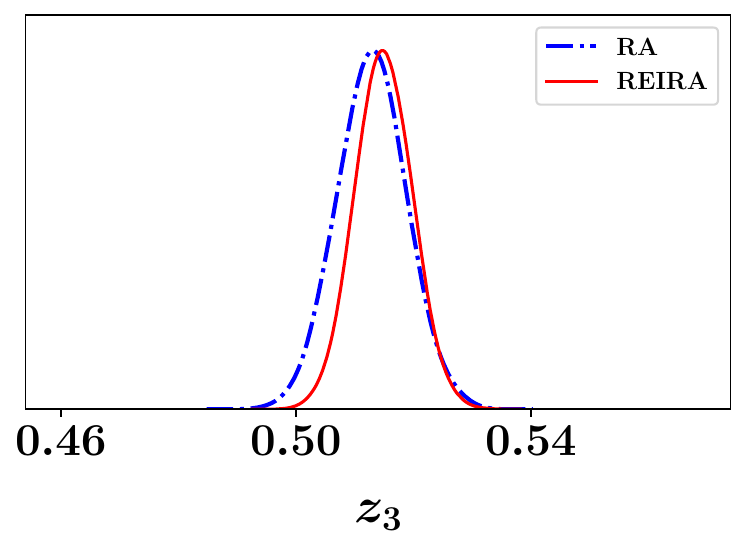}
    \includegraphics[width=0.2235\linewidth]{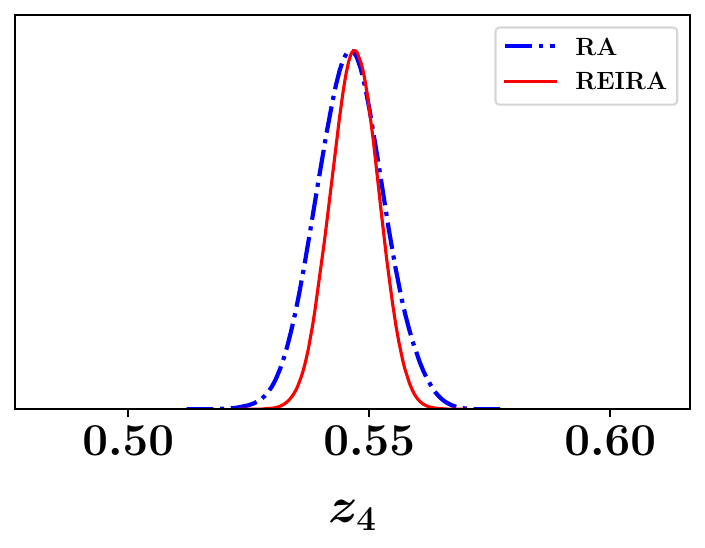}
    \includegraphics[width=0.23\linewidth]{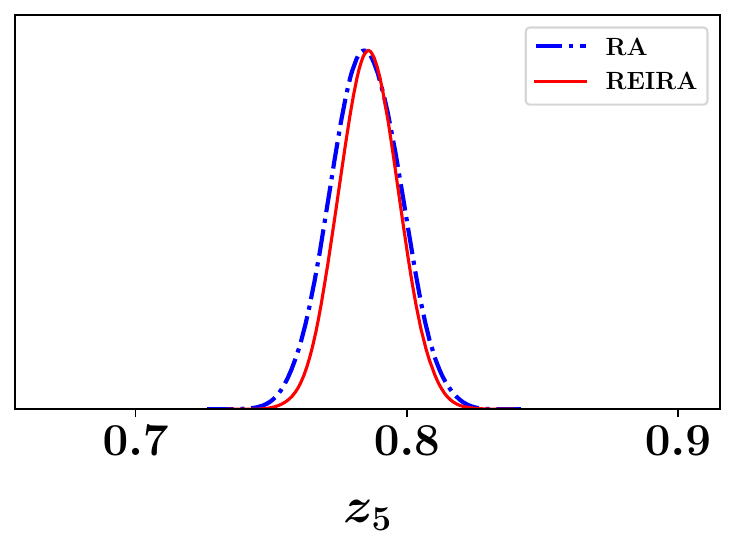}
    \includegraphics[width=0.23\linewidth]{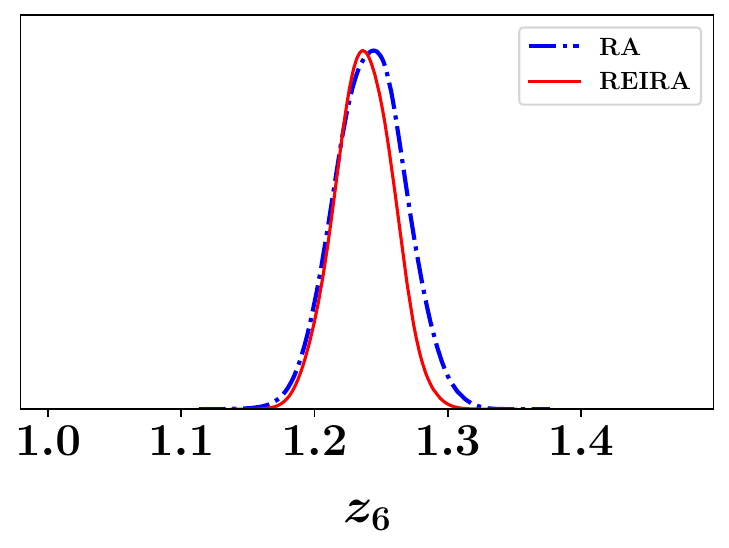}
    \includegraphics[width=0.23\linewidth]{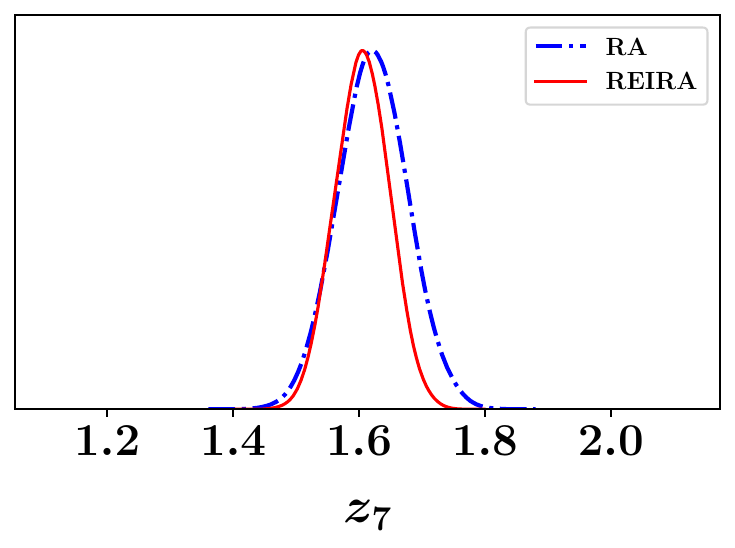}
    \caption{Comparison of the uncertainty in characteristic redshifts obtained using RA (blue) and REIRA (red).}
    \label{fig:z_knots_rc_vs_no_rc}
\end{figure*}

\begin{figure*}[t]
    \centering
     \begin{subfigure}[b]{0.75\textwidth}
        \includegraphics[width=\textwidth]{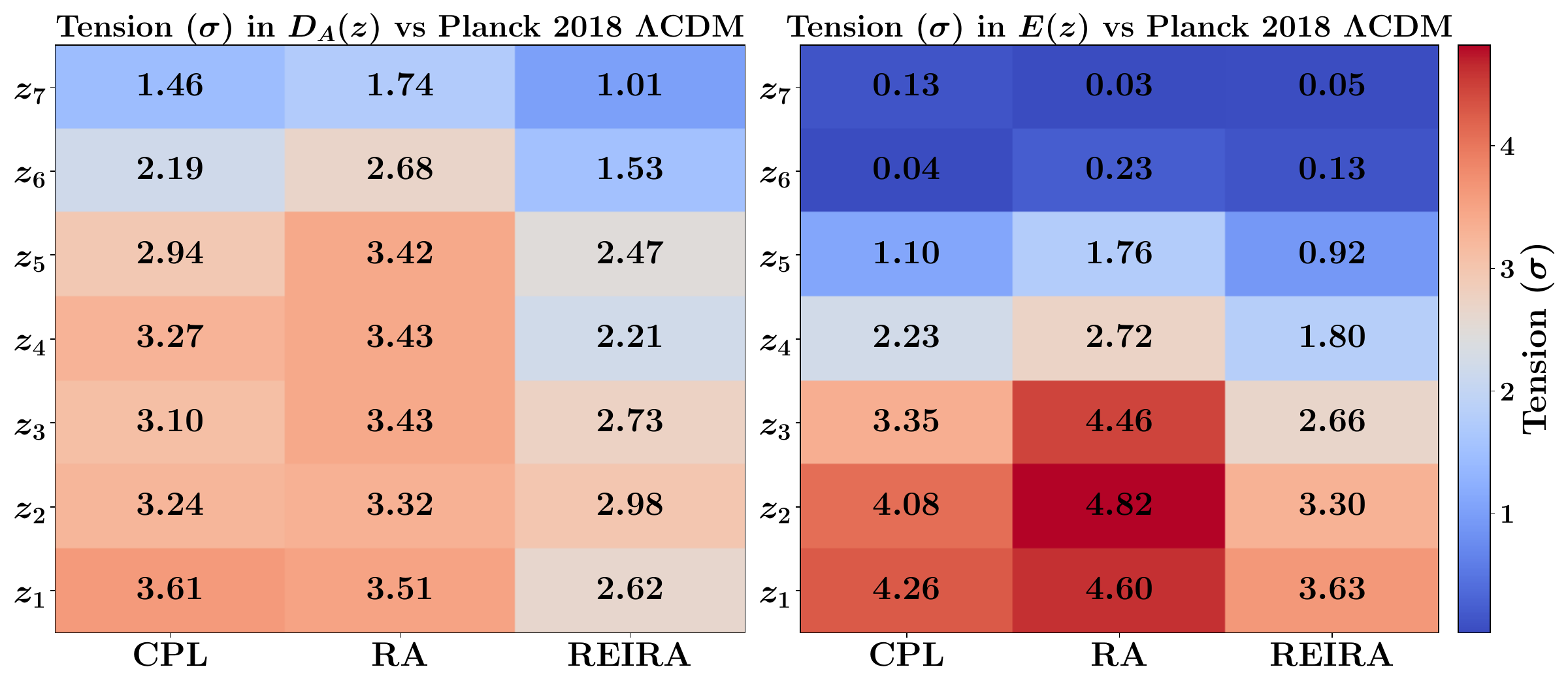}
        \caption{}
        \label{fig5:sub1}
    \end{subfigure}
    \begin{subfigure}[b]{0.5\textwidth}
        \includegraphics[width=\textwidth]{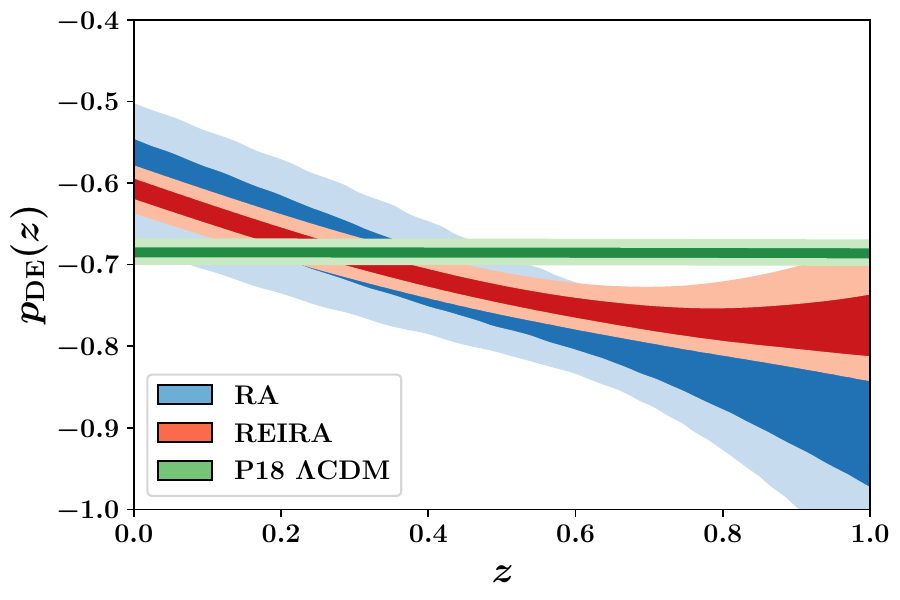}
        \caption{}
        \label{fig5:sub2}
    \end{subfigure}
       \begin{subfigure}[b]{0.45\textwidth}
        \includegraphics[width=\textwidth]{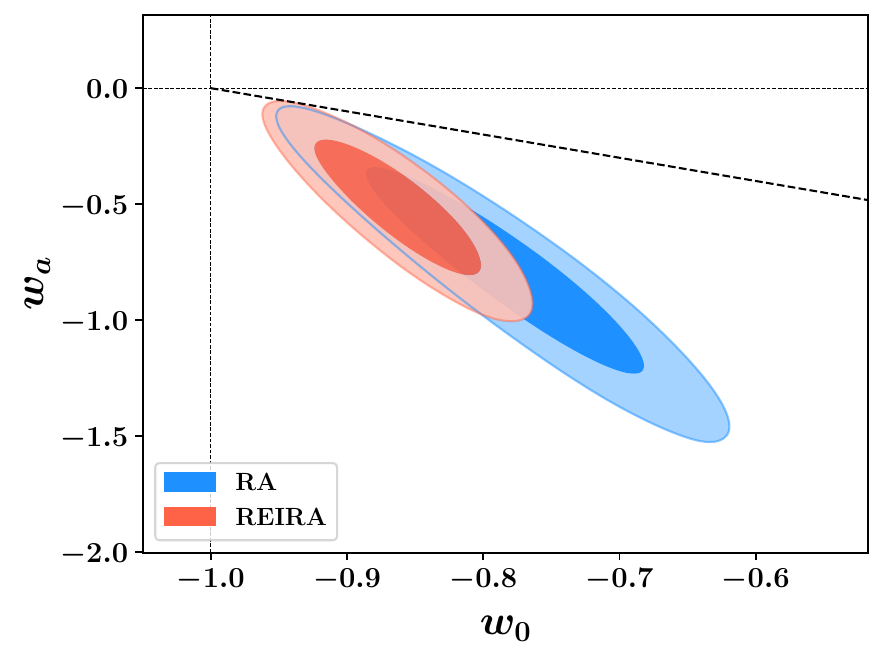}
        \caption{}
        \label{fig5:sub3}
    \end{subfigure}
    \caption{Heatmap comparison of the statistical tensions in $D_A$ and $E$ with respect to the $\Lambda$CDM baseline, is shown for three cases: RA, REIRA, and the CPL model in \ref{fig5:sub1}. Blue shades denote lower tension, while red shades denote higher tension; lighter and deeper tones within each color represent finer gradations. Evolution of the DE pressure $p_{\rm DE}(z)$ obtained using RA (blue) and REIRA (red), compared to the $\Lambda$CDM baseline (green) is shown in \ref{fig5:sub2}. Constraints in the $w_0$–$w_a$ parameter space for the CPL model, obtained from compressed likelihoods based on RA (blue) and REIRA (red) are shown in \ref{fig5:sub3}. Deeper and lighter shades in \ref{fig5:sub2} and \ref{fig5:sub3} correspond to $1\sigma$ and $2\sigma$ confidence levels respectively.}
    \label{}
    \end{figure*}

\section{Raychaudhuri Equation Informed Reconstruction Algorithm} \label{sec:rc}

The reconstruction results discussed so far are obtained solely from observational datasets, without incorporating any physics-based constraints. In this section, we investigate how these results improve when essential physical information is included in the reconstruction algorithm. For this purpose, we employ the RE as a physics prior.

The RE for a timelike congruence having velocity $u^\mu$ ($u_\mu u^\mu = -1$) is given by  \citep{Raychaudhuri:1953yv, Ehlers1993, Kar2007, Ellis2007}
\begin{equation}\label{rceq}
\begin{split}
  \frac{ \mathrm{d}{{\theta}}}{{\mathrm{d}\tau} }= - \frac{1}{3} {\theta}^2 - {\sigma}_{\mu\nu}{\sigma}^{\mu\nu}
 +{\omega}_{\mu\nu}{\omega}^{\mu\nu} +\nabla_\mu A^\mu - {{R}}_{\mu\nu} u^{\mu} u^{\nu}\,,
 \end{split}
\end{equation}
where ${\theta}={\nabla}_\mu u^\mu$ is the expansion scalar, $\tau$ is proper time, 
${\sigma}_{\mu\nu}={\nabla}_{(\nu}u_{\mu)}-\frac{1}{3}h_{\mu\nu}{\theta}$
is the shear tensor with $h_{\mu\nu} = g_{\mu\nu} + u_{\mu}u_{\nu}$ being the induced spatial metric, ${\omega}_{\mu\nu}={\nabla}_{[\nu}u_{\mu]}$
is the rotation tensor, $A^\mu=u^\alpha \nabla_\alpha u^\mu$ is the four acceleration and ${{R}}_{\mu\nu}$ is the Ricci tensor. The RE is a geometric identity in Riemannian geometry and becomes a kinematic one when we use the Einstein's equation. 

For a spatially flat FLRW geometry, the RE, in terms of redshift $z$, can simply be written as
\begin{equation}
    2(1+z)EE^\prime - 3E^2 = \frac{P}{3H_0^2}\equiv p_{\rm DE},
\end{equation}
where $P$ is the total pressure of the Universe, with $8\pi G = 1$. At late times, only DE contributes to the pressure (given that matter is pressureless, and there is negligible contribution from radiation at late times), and this must be negative in order to drive accelerated expansion. Translating this condition into the comoving distance $D_M$, we obtain
\begin{equation}
    2(1+z)\frac{D_M^{\prime\prime}}{{D_M^\prime}^3}+\frac{3}{{D_M^\prime}^2}
    = -p_{\rm DE} > 0.
\end{equation}
We impose this inequality within the range $0 < z < 3$ in the knot-based reconstruction. We stress that this prior holds true irrespective of whether DE is non-interacting or interacting or whether it is canonical or non-canonical. With this, we introduce, for the first time, \textbf{Raychaudhuri Equation Informed Reconstruction Algorithm (REIRA)}. 

The effect of REIRA  on the characteristic redshifts is shown in Fig.~\ref{fig:z_knots_rc_vs_no_rc}. As expected from their robustness, the mean values of these redshifts do not change significantly compared to RA. However, the error bars shrink once the RE constraint is enforced, with the reduction being particularly significant at higher redshifts\footnote{This can also be seen in the comparison between Tables~\ref{tab:spline4} and \ref{tab:recon-rc} in Appendix where the latter summarizes the reconstructed values of $D_A(z)$ and $E(z)$ for REIRA.}.

Figure~\ref{fig4:sub2} shows the residual $\Delta H(z)$ for REIRA and the CPL model compared with the $\Lambda$CDM baseline. Similar to the RA case (Fig.~\ref{fig4:sub1}), REIRA remains close to CPL, however the mean values of $\Delta H$ decrease and the error bars get reduced as compared to the RA case. This demonstrates that incorporating the REIRA substantially enhances the constraining power of reconstruction methods. We should emphasize that since the RE is fundamentally a kinematic identity, REIRA does not compromise the model independence of the reconstruction framework. 

We examine the impact of REIRA on the inferred tensions with respect to the $\Lambda$CDM baseline. To provide a clear comparison, we present in Fig.~\ref{fig5:sub1} a heatmap illustrating the tensions in $D_A$ and $E$ for three cases: RA, REIRA, and CPL model\footnote{The corresponding tensions in the quantities $D_A$ and $E$ for the cases of RA, CPL and REIRA are summarized using Tables \ref{tab:spline4}, \ref{tab:cpl} and \ref{tab:recon-rc} respectively in the Appendix section.}. As evident from the figure, REIRA reduces the tensions relative to $\Lambda$CDM but the tensions in $E(z)$ at low redshifts still remain at more than $2.5-3.5\sigma$ confirming significant deviation from $\Lambda$CDM.

To further highlight the impact of the REIRA, we show the reconstructed evolution of the DE pressure $p_{\rm DE}(z)$ in Fig.~\ref{fig5:sub2}. The REIRA yields significantly tighter bounds across the redshift range and displays a smoother, more stable trend at higher redshifts. This improvement arises because the RE directly links the pressure to the observable $D_M(z)$. The reconstructed DE pressure for REIRA clearly shows a significant departure from a constant pressure in $\Lambda$CDM case.

\section{Compressed Likelihood from Reconstructed Data Vectors} \label{sec:like}

Building on the robustness of the characteristic redshifts, we construct compressed likelihoods from the corresponding data vectors obtained from both RA and REIRA. 

To generate the compressed data vectors, we utilize the normalized Hubble parameter $E(z_i) \equiv H(z_i)/H_0$ values along with their uncertainties, taking into account two sources of error: reconstruction errors in $D_M(z_i)$ and the finite precision in the determination of $z_i$'s. Correlations between different $z_i$, induced by the reconstruction method, are included in the full covariance matrix $\mathcal{C}_{ij}$, ensuring that the compressed vector $\mathbf{d}_{\rm vec} = \left\{ E(z_1), E(z_2), \dots, E(z_7) \right\}$ accurately captures the essential constraining information from the full dataset while significantly reducing its dimensionality.

The $\mathbf{d}_{\rm vec}$ is then combined with the compressed CMB likelihood \citep{Wang:2007mza, Bansal:2025ipo} for an efficient exploration of the DE parameter space. As a demonstrative example, we adopt the CPL parametrization, $w(z) = w_0 + w_a \frac{z}{1+z}$, and examine the resulting constraints in the $w_0$–$w_a$ plane. 

In Fig.~\ref{fig5:sub3}, we compare the constraints in the $w_0$–$w_a$ parameter space obtained from RA and REIRA. The latter leads to a substantial tightening of the allowed parameter region, \emph{highlighting the gain in constraining power when physics-based information is incorporated into the reconstruction framework}.

\section{Conclusion} \label{sec:conclusion}

Here we summarize the novel aspects of this paper:
\begin{itemize}
\item 
We show that the previously obtained characteristic redshifts by \citet{Mukherjee:2025ytj} using geometrical properties of various cosmological distances are stable under different DE parameterizations as well as reconstruction algorithms, making them reliable and robust geometric anchors of the expansion history.

\item 
We further show that the Alcock–Paczyński corrections at these redshift anchors are found to be unity and hence fiducial model dependence is minimized. This reaffirms their potential as powerful and reliable probes for exploring possible new physics.

\item 
We also show that at three low redshift anchors, the deviations in expansion rate from Planck-$\Lambda$CDM model are significantly larger irrespective of the reconstruction algorithm or the DE parametrization like CPL. This confirms robust hints for new physics at low redshift cosmological evolutions. 

\item 
Finally, for the first time, we introduce the REIRA (\textbf{Raychaudhuri Equation Informed Reconstruction Algorithm}) for model independent reconstruction of the cosmological evolution using low-redshift data. We show that REIRA significantly improves the reconstruction method resulting in much tighter constraints on cosmological evolution as well as DE parameter space.

\end{itemize}

This further motivates us to construct {\bf Raychaudhuri Equation Informed Neural Network Algorithm (REINN) \citep{reinn}} for cosmological reconstruction which will be our next goal.

\begin{acknowledgments}
SGC acknowledges funding from the Anusandhan National Research Foundation (ANRF), Govt. of India, under the National Post-Doctoral Fellowship (File no. PDF/2023/002066). PM acknowledges funding from ANRF, Govt. of India, under the National Post-Doctoral Fellowship (File no. PDF/2023/001986). AAS acknowledges the funding from ANRF, Govt. of India, under the research grant no. CRG/2023/003984. We acknowledge the use of the HPC facility, Pegasus, at IUCAA, Pune, India. This article/publication is based upon work from COST Action CA21136- ``Addressing observational tensions in cosmology with systematics and fundamental physics (CosmoVerse)'', supported by COST (European Cooperation in Science and Technology).
\end{acknowledgments}

\software{numpy \citep{2020NumPy-Array},
          scipy \citep{2020SciPy-NMeth}, 
          matplotlib \citep{Hunter:2007},
          emcee \citep{Foreman_Mackey_2013},
          GetDist \citep{Lewis:2019xzd},
          fgivenx \citep{fgivenx}.
          }

\bibliography{ref}{}
\bibliographystyle{aasjournalv7}

\appendix

\begin{table*}[h]
\centering
\caption{Reconstructed values of $D_A(z)$ and $E(z)$ at the characteristic redshifts $z_i$, obtained using RA. The Planck 2018 $\Lambda$CDM predictions are also shown, along with the tensions in units of $\sigma$.}
\renewcommand{\arraystretch}{1.25}
\setlength{\tabcolsep}{6pt}
\begin{tabular}{ccccccc}
\toprule
$z_i$ & $D_A(z_i)$ & $D_A^{\text{P18}}(z_i)$ & Tension ($\sigma$) 
      & $E(z_i)$ & $E^{\text{P18}}(z_i)$ & Tension ($\sigma$) \\
\hline
$z_1$=0.350 $\pm$ 0.003 &  0.232 $\pm$ 0.001  & 0.237 $\pm$ 0.001 & 3.23 & 1.246 $\pm$ 0.007  & 1.208 $\pm$ 0.005 & 4.52 \\
$z_2$=0.414 $\pm$ 0.004 &  0.257 $\pm$ 0.001  & 0.263 $\pm$ 0.002 & 3.34 & 1.296 $\pm$ 0.006  & 1.256 $\pm$ 0.006 & 4.77 \\
$z_3$=0.513 $\pm$ 0.006 &  0.289 $\pm$ 0.001  & 0.296 $\pm$ 0.002 & 3.34 & 1.375 $\pm$ 0.004  & 1.333 $\pm$ 0.009 & 4.34 \\
$z_4$=0.546 $\pm$ 0.007 &  0.299 $\pm$ 0.001  & 0.306 $\pm$ 0.002 & 3.39 & 1.402 $\pm$ 0.013  & 1.360 $\pm$ 0.009 & 2.67 \\
$z_5$=0.785 $\pm$ 0.014 &  0.348 $\pm$ 0.001  & 0.356 $\pm$ 0.003 & 3.34 & 1.612 $\pm$ 0.013  & 1.574 $\pm$ 0.017 & 1.74 \\
$z_6$=1.242 $\pm$ 0.027 &  0.388 $\pm$ 0.002  & 0.397 $\pm$ 0.002 & 2.64  & 2.073 $\pm$ 0.046  & 2.059 $\pm$ 0.036 & 0.24 \\
$z_7$=1.622 $\pm$ 0.056 &  0.396 $\pm$ 0.004  & 0.403 $\pm$ 0.002 & 1.73 & 2.528 $\pm$ 0.024  & 2.524 $\pm$ 0.077 & 0.05 \\
\hline
\end{tabular} \label{tab:spline4}
\end{table*}

\begin{table*}[h]
\centering
\caption{Values of $D_A(z)$ and $E(z)$ at the characteristic redshifts $z_i$ for the CPL model, obtained from the MCMC chains. The Planck 2018 $\Lambda$CDM predictions are also shown, along with the tensions in units of $\sigma$.}
\renewcommand{\arraystretch}{1.25}
\setlength{\tabcolsep}{6pt}
\begin{tabular}{ccccccc}
\toprule
$z_i$ & $D_A(z_i)$ & $D_A^{\text{P18}}(z_i)$ & Tension ($\sigma$) 
      & $E(z_i)$ & $E^{\text{P18}}(z_i)$ & Tension ($\sigma$) \\
\hline
$z_1$=0.350 $\pm$ 0.003 &  0.232 $\pm$ 0.001  & 0.237 $\pm$ 0.002 & 3.24 & 1.245 $\pm$ 0.007  & 1.209 $\pm$ 0.005 & 4.26 \\
$z_2$=0.415 $\pm$ 0.005 &  0.258 $\pm$ 0.001  & 0.264 $\pm$ 0.002 & 3.1 & 1.293 $\pm$ 0.006  & 1.257 $\pm$ 0.007 & 4.08 \\
$z_3$=0.516 $\pm$ 0.007 &  0.290 $\pm$ 0.001  & 0.297 $\pm$ 0.002 & 2.94 & 1.369 $\pm$ 0.004  & 1.335 $\pm$ 0.009 & 3.35 \\
$z_4$=0.548 $\pm$ 0.006 &  0.299 $\pm$ 0.001  & 0.306 $\pm$ 0.002 & 3.61 & 1.394 $\pm$ 0.012  & 1.362 $\pm$ 0.009 & 2.23 \\
$z_5$=0.789 $\pm$ 0.012 &  0.349 $\pm$ 0.001  & 0.357 $\pm$ 0.002 & 3.27 & 1.600 $\pm$ 0.011  & 1.579 $\pm$ 0.016 & 1.1 \\
$z_6$=1.243 $\pm$ 0.019 &  0.390 $\pm$ 0.002  & 0.397 $\pm$ 0.002 & 2.19  & 2.062 $\pm$ 0.034  & 2.060 $\pm$ 0.029 & 0.04 \\
$z_7$=1.622 $\pm$ 0.037 &  0.397 $\pm$ 0.003  & 0.403 $\pm$ 0.002 & 1.46 & 2.517 $\pm$ 0.021  & 2.525 $\pm$ 0.054 & 0.13 \\
\hline
\end{tabular} \label{tab:cpl}
\end{table*}

\begin{table*}[h]
\centering
\caption{Reconstructed values of $D_A(z)$ and $E(z)$ at the characteristic redshifts $z_i$, obtained from REIRA. The Planck 2018 $\Lambda$CDM predictions are also shown, along with the tensions in units of $\sigma$.}
\renewcommand{\arraystretch}{1.25}
\setlength{\tabcolsep}{6pt}
\begin{tabular}{ccccccc}
\toprule
$z_i$ & $D_A(z_i)$ & $D_A^{\text{P18}}(z_i)$ & Tension ($\sigma$) 
      & $E(z_i)$ & $E^{\text{P18}}(z_i)$ & Tension ($\sigma$) \\
\hline
$z_1$=0.351 $\pm$ 0.002 &  0.235 $\pm$ 0.001  & 0.237 $\pm$ 0.001 & 2.98 & 1.231 $\pm$ 0.004  & 1.209 $\pm$ 0.005 & 3.63 \\
$z_2$=0.416 $\pm$ 0.003 &  0.260 $\pm$ 0.001  & 0.264 $\pm$ 0.001 & 2.73 & 1.279 $\pm$ 0.003  & 1.257 $\pm$ 0.006 & 3.3 \\
$z_3$=0.515 $\pm$ 0.005 &  0.293 $\pm$ 0.001  & 0.297 $\pm$ 0.002 & 2.47 & 1.357 $\pm$ 0.003  & 1.335 $\pm$ 0.008 & 2.66 \\
$z_4$=0.547 $\pm$ 0.005 &  0.302 $\pm$ 0.001  & 0.306 $\pm$ 0.002 & 2.62 & 1.383 $\pm$ 0.009  & 1.361 $\pm$ 0.008 & 1.8 \\
$z_5$=0.786 $\pm$ 0.011 &  0.352 $\pm$ 0.001  & 0.357 $\pm$ 0.002 & 2.21 & 1.592 $\pm$ 0.010  & 1.575 $\pm$ 0.015 & 0.92 \\
$z_6$=1.237 $\pm$ 0.022 &  0.392 $\pm$ 0.002  & 0.397 $\pm$ 0.002 & 1.53 & 2.060 $\pm$ 0.039  & 2.053 $\pm$ 0.032 & 0.13 \\
$z_7$=1.605 $\pm$ 0.043 &  0.399 $\pm$ 0.003  & 0.403 $\pm$ 0.002 & 1.01 & 2.505 $\pm$ 0.021  & 2.502 $\pm$ 0.060 & 0.05 \\
\hline
\end{tabular} \label{tab:recon-rc}
\end{table*}

\end{document}